\documentclass[authoryear]{FLO_v1}%

\usepackage{graphicx}
\usepackage{upgreek}
\usepackage{multicol,multirow}
\usepackage{amsmath,amssymb,amsfonts}
\usepackage{mathrsfs}
\usepackage{amsthm}
\usepackage[figuresright]{rotating}
\usepackage{appendix}
\usepackage[authoryear]{natbib}
\usepackage{ifpdf}
\usepackage[T1]{fontenc}
\usepackage{newtxtext}
\usepackage{newtxmath}
\usepackage{textcomp}
\usepackage{xcolor}

\usepackage[colorlinks,allcolors=blue]{hyperref}
\definecolor{jourcolor}{cmyk}{1,0.57,0.01,0.38}
\hypersetup{
    colorlinks,%
    citecolor=jourcolor,%
    filecolor=jourcolor,%
    linkcolor=jourcolor,%
    urlcolor=jourcolor
}

\theoremstyle{definition}

\articletype{RESEARCH ARTICLE}

\DOI{}

\Year{2021}

\Vol{This draft was prepared using the LaTeX style file belonging to FLOW: Applications of Fluid Mechanics}

\Price{}

\art-id{}

\citearticle{N. S. Lagopoulos, G. D. Weymouth and B. Ganapathisubramani}

\begin{document}

\title[AR effect on tandem flapping foils]{Effect of aspect ratio on the propulsive performance of tandem flapping foils}

\author{N. S. Lagopoulos $^{1,3\ast}$, G. D. Weymouth $^{2,4\ast}$, B. Ganapathisubramani $^1$}

\address[1]{ Aerodynamics and Flight Mechanics Group, University of Southampton, UK}
\address[2]{ Southampton Marine and Maritime Institute, University of Southampton, UK}
\address[3]{Dolprop Industries AB, Ekerö, Sweden}
\address[4]{Alan Turing Institute, London, UK}

\corres{*}{Corresponding authors. E-mails:
\emaillink{nikolaos@dolprop.se} and \emaillink{g.d.weymouth@soton.ac.uk}}

\keywords{Vortex dynamics; Swimming/flying; Autonomous underwater vehicles}

\date{\textbf{Received:} XX 2021; \textbf{Revised:} XX XX 2021; \textbf{Accepted:} XX XX 2021}

\abstract{In this work, we describe the impact of aspect ratio (AR) on the performance of optimally phased, identical flapping flippers in a tandem configuration. Three-dimensional simulations are performed for seven sets of single and tandem finite foils at a moderate Reynolds number, with thrust producing, heave-to-pitch coupled kinematics. Increasing slenderness (or aspect ratio - AR) is found to improve thrust coefficients and thrust augmentation but the benefits level off towards higher values of AR. On the other hand, the propulsive efficiency shows no significant change with increasing AR, while the hind foil outperforms the single by a small margin. Further analysis of the spanwise development and propagation of vortical structures allows us to gain some insights on the mechanisms of these wake interactions and provide valuable information for the design of novel biomimetic propulsion systems. }

\maketitle

\begin{boxtext}

\textbf{\mathversion{bold}Impact Statement}
Tandem flapping foils has the potential to be used for propulsion, especially among bio-inspired AUV designers, due to their superior performance over single flippers. In this study, we evaluate the importance of aspect ratio on the thrust-augmenting effect of in-line flapping, known as wake recapture. It is shown that flipper elongation impacts the interaction between the hind flipper (or follower) and its incoming flow, as it strengthens the trailing edge vortices, shed in the wake of the front flipper. This affects both the thrust generating capacity and the optimal phasing of the flippers, allowing the engineer to determine the vehicle´s suitability towards certain missions, simply based on foil slenderness. An in-depth analysis of the wake dynamics enables us to distinguish the limitations as well as ways to optimize this approach by monitoring the transition towards a quasi two-dimensional flow.

\end{boxtext}
\section{Introduction} \label{Intro}
Flapping foil mechanisms are the basic means of propulsion and control within the avian and aquatic fauna. These systems are often more agile, durable and efficient compared to conventional man-made propulsors \citep{Weymouth2016}. Thus, many studies have focused on the analysis of these biological configurations in terms of kinematics \citep{Khalid2021,Cimarelli2021} fluid-structure interaction \citep{Kim2013,Zurman2020} as well as the effects of planform geometry \citep{Dagenais2020,Zurman2021} and flexibility \citep{Shi2020,Fernandez2021}.

\begin{figure} 
\centerline{\includegraphics[width =4.5 in]{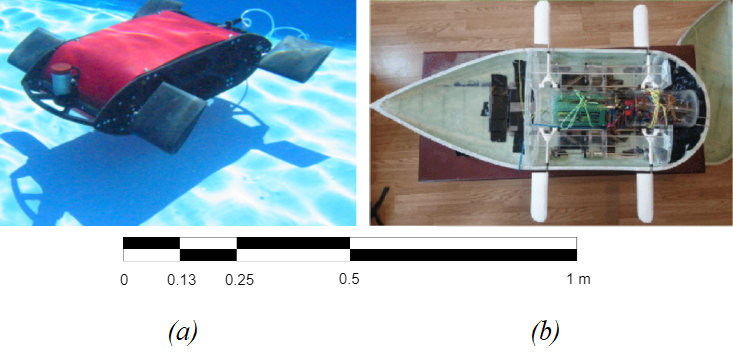}}
\caption{Two examples of bio-inspired AUVs that combine front and back flipper oscillation (tandem arrangement) as a means of propulsion: (a) performs a pure pitching motion \citep{Long2006} while (b) uses a combination of rolling and pitching \citep{Weymouth2017}. A simplified version of the latter kinematics is utilized by this study.}
\label{agr}
\end{figure}

Tandem flapping configurations e.g. insect wings \citep{Alexander1984,Thomas2004}, plesiosaur flippers \citep{Robinson1975,Hawthorne2019} etc. are shown to outperform single flappers due to certain foil-wake interactions commonly referred to as $wake ~recapture$ \citep{Broering2012,Muscutt2017}. This has inspired researchers to experiment on quadruple foil systems to propel autonomous underwater vehicles (AUVs), utilizing a variety of harmonic kinematics e.g pitch, roll, coupled motion etc. Most of these tetrapodal swimmers (see figure \ref{agr}) are electric-powered, designed for a wide range of depths ($1 m-100 m$) and can reach velocities of $0.5m/s -2 m/s$ \citep{Licht2004, Long2006, Weymouth2017}, which are comparable to modern propeller-driven, ocean-going AUVs of a similar size and weight \citep{Yuh2000}.

Of particular interest, towards the design of these systems, is the $aspect$ $ratio$ AR (for rectangular wings AR $=W/\mathcal{C}$ where $W$ is the wingspan and $\mathcal{C}$ is the chord length) due to its impact on the system´s thrust generation capacity. Slender flippers for example, are widely considered beneficial to both thrust and efficiency \citep{Green2008,Shao2010,Dewey2013} which is further implied by the predominance of high AR wings among birds and insects \citep{Ellington1984,Azuma1992,Usherwood2002}. As a result, early research was driven towards two dimensional or quasi two dimensional approaches both experimentally \citep{Koochesfahani1989,Triantafyllou1993} and numerically \citep{Pedro2003,Mittal2004,Guglielmini2004}.

Unlike avian organisms however, aquatic animals demonstrate a great variety of AR with non migratory fish utilizing mostly low AR \citep{Walker2002,Combes2001} as they are considered more suitable for their drag-based paddling motion. Furthermore, comparisons among various species  suggest that high AR‘s benefit cruising efficiency  while low AR‘s promotes thrust generation in short bursts \citep{Flammang2009,Domenici2010} which is also supported by recent experiments \citep{Lee2017}. A preference towards lower AR in aquatic propulsion can be additionally attributed to the much higher density of water, which leads to greater added-mass associated bending moments \citep{Dong2006}. This can significantly constrain the design of an AUV by determining manufacturing costs, durability, mission envelope etc. and thus demonstrates the necessity of finite flipper analysis.

Contemporary literature on finite wings of varying AR often focuses on single flapping configurations  \citep{Zurman2021,Hammer2021,Zhong2021} with only a few studies related to tandem arrangements \citep{Arranz2020}. A key feature of the above is the presence of tip vortices that transform the two dimensional wake into a complex chain of ring-like formations \citep{Shao2010,Li2018}. Moreover, the majority of these studies utilizes insect and small fish kinematics and/or geometries which, although quite suitable for special applications, are less relevant to open water designs. On the other hand, certain heave-to-pitch combinations are considered dominant in cetacean locomotion  \citep{Ayancik2020,Han2020} (where spanwise flexibility of the caudal fin is comparatively low \citep{Gough2018,Adams2019}). Furthermore, heave-to-pitch coupling is considered sufficient to represent the mid-chord kinematics of flipper-based, AUV´s and/or aquatic animals using roll-to-pitch combinations \citep{MuscuttEX2017} such as sea turtles, penguins and most notably the tetrapodal plesiosaurs \citep{Carpenter2010}. Besides, the effect of flipper AR on the wake recapture remains unknown, despite its aforementioned importance within tandem flapping AUV consepts. 

The present study attempts to address these issues via the numerical analysis of rectangular flippers with elliptical tip, undergoing heave-to-pitch coupling for a chord based Reynolds number, $Re_\mathcal{C}= \rho U_{\infty} \mathcal{C} / \mu =8500$ where $\rho$ is the water density, $U_{\infty}$ is the freestream velocity and $\mu$ is the dynamic viscosity. Seven AR are tested in both single and tandem configurations of identical AR for an amplitude based Strouhal, $St_{A}= f\cdot 2\mathcal{A}/ U_{\infty}=0.4 $ where $f$ is the frequency of oscillation and $2\mathcal{A}$ is the peak-to-peak amplitude of the trailing edge (TE) \citep{Triantafyllou1991}. In addition, the phase lag and distance between consecutive flippers are kept constant, selected for maximum thrust augmentation at the given $St_A$ in 2D \citep{Muscutt2017}. Here, the choice of $St_A$ is based on the observed range of Strouhals utilized by swimming and flying organisms \citep{Triantafyllou1993}. Furthermore, the test cases are evaluated in terms of thrust coefficient, relative thrust augmentation and hydrodynamic efficiency. To this end, we compare the single/tandem flipper sensitivity to AR and attempt to shed light on the three dimensional aspect of the wake to wake interaction.

\section{Methodology}
\subsection{Flipper Geometry and kinematics}

\begin{figure}
  \centerline{\includegraphics[width =4 in]{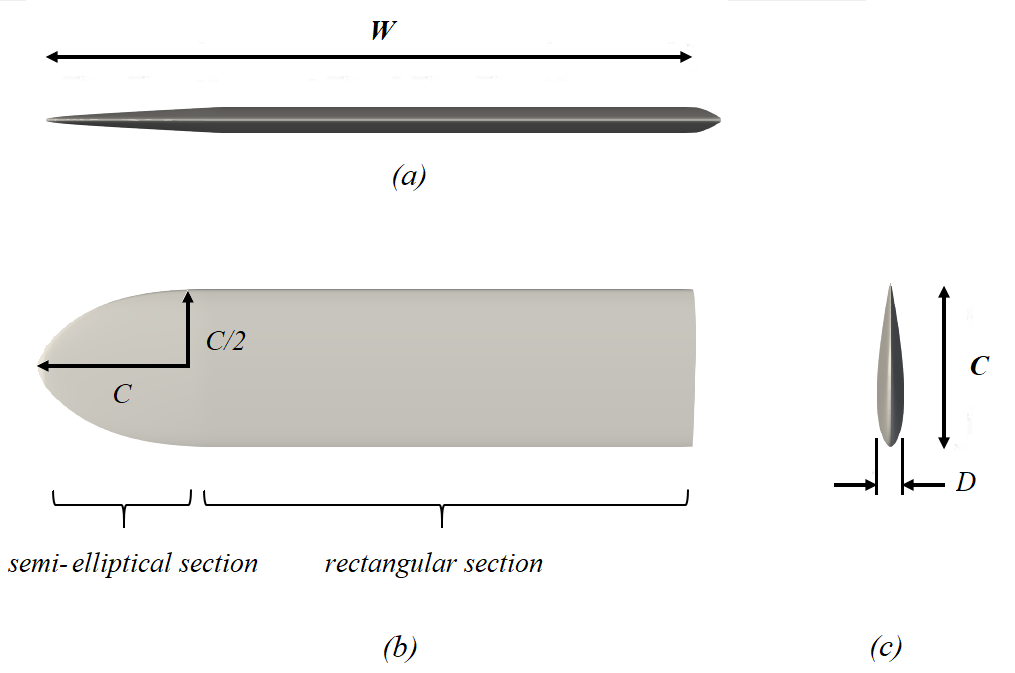}}
  \caption{Structural details of an AR=4 hydrofoil, where the (a) frontal,(b) upper and (c) side view are presented. A detailed model of the flipper in the form of an IGS file can be found online, within the supplementary material of this study.}\label{flipperx}
\end{figure}

We consider a rigid NACA0016 with a thickness $D = 0.16~\mathcal{C}$, a rectangular planform section where the width is equal to $1~\mathcal{C}$ and a tapered elliptical tip as shown in figure \ref{flipperx}. Here the elliptical section has a span of $1~\mathcal{C}$ while $W$ is the total span of the flipper. Thus, for the sake of simplicity, we use the AR definition of rectangular flippers (explained in section \ref{Intro}) and we set our baseline test case at AR $=2$ proceeding towards AR $=8$ in increments of AR $=1$.

\begin{figure}
  \centerline{\includegraphics[width =4 in]{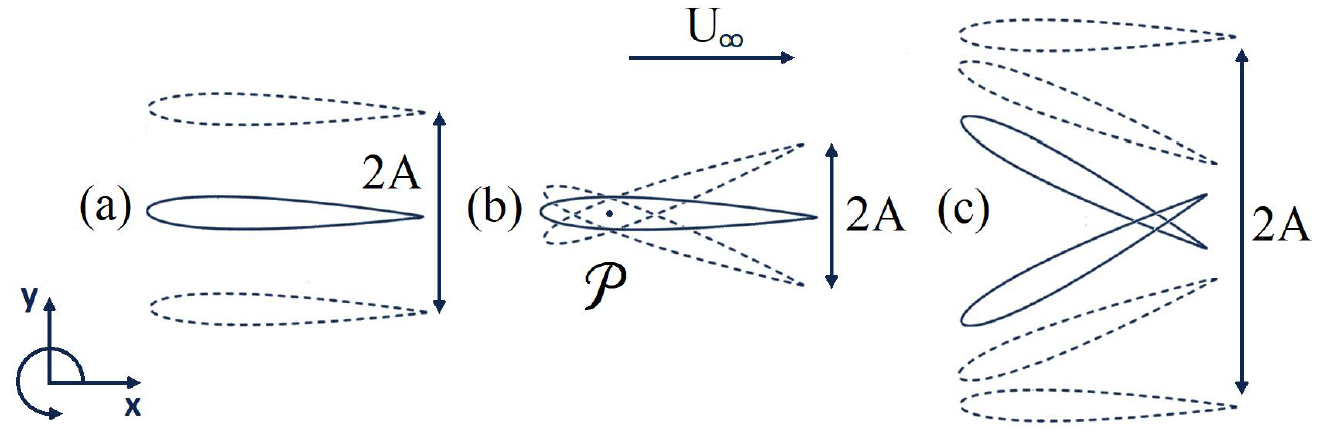}}
  \caption{The kinematic parameters and coordinate system of an oscillating foil undergoing (a) heave (b) pitch and (c) coupled motion. Redrawn from \citep{Lagopoulos2019}.}
\label{fig:2o}
\end{figure}

The kinematics parameters of the hydrofoil can be seen in figure \ref{fig:2o}. As stated above, the flippers utilize heave to pitch coupling which is achieved by the superposition of the two harmonic components. More specifically, pitch refers to the sinusoidal rotation about the pivot point $\mathcal{P} = 0.25$ (normalised by $\mathcal{C}$) while heave is a sinusoidal, vertical translation with respect to the centreline: 
\begin{eqnarray}
& y_f(t) =\underbrace{h_\mathit{0}\sin(2f\pi t)}_{y_h(t)} + \underbrace{(1-\mathcal{P})\mathcal{C}\sin[{\theta}(t)]}_{y_\theta(t)}\\
& with~~~~~~ \theta(t) = \theta_\mathit{0} \sin(2f\pi t +\psi)
\label{eq:xdef}
\end{eqnarray}
where subscripts $f$, $h$ and $\theta$ denote the front (or single foil), heaving and pitching components respectively.

Here, the instantaneous pitching angle is expressed as the $\theta(t)$, while $h_{0}$ and $\theta_{0}$ represent the amplitudes of the two motions. Note that although the total peak-to-peak amplitude is a combination of these TE displacements, the chosen kinematic parameters result in $\mathcal{A} \sim  h_{0} = 1~\mathcal{C}$. Furthermore, the heave to pitch phase difference is set to $\psi$ = $90^{\circ}$, which is shown to maximize the propulsive efficiency within the frequency range of interest \citep{Platzer2008}. Lastly, the combined (or effective) angle of attack $\alpha(t)$ equals to the summation of $\theta(t)$ and the heave-induced angle of attack so that its amplitude is expressed as:

\begin{equation}
\alpha_0 = \arctan \frac{2 \pi f h_{0}}{U_\infty} - \theta_{0}
\end{equation}
where $2 \pi f h_{0}$ is the amplitude of $dy_h/dt$. Within this study, all simulations are conducted for $\alpha_0=20^{\circ}$ due to its dominance within modern cetaceans \citep{Fish1999} and relevant studies in tetrapodal swimming \citep{MuscuttEX2017}.

\begin{figure}
  \centerline{\includegraphics[width =3.3 in]{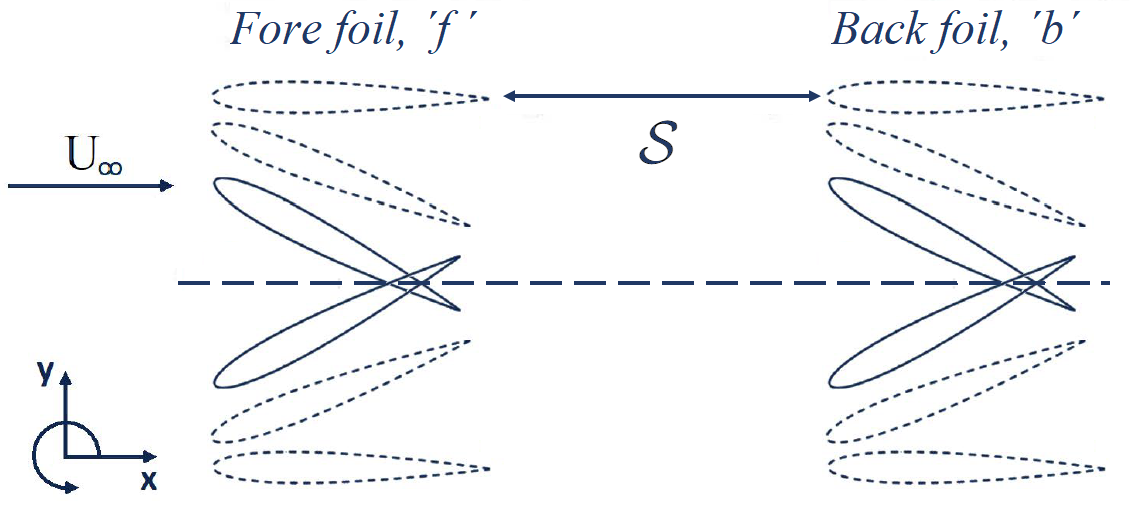}}
  \caption{Details of a tandem foil configuration, undergoing synchronous ($\phi=0^\circ$), heave-to-pitch coupling.}\label{fig:3o}
\end{figure}

The complete tandem arrangement is depicted in figure \ref{fig:3o}. To distinguish parameters referring to the front or the back foil, we utilize the subscripts $f$ (already mentioned above) and $b$ respectively, for the remaining of this study. Furthermore, to describe the foil-to-foil interaction, we introduce two more parameters: the $\emph{phase}~\emph{lag}$ and the inter foil $\emph{spacing}$. 

The phase lag between the two foils is expressed as $\varphi$ and will be referred to as simply the $phase$. Thus the back flipper´s motion is described as: 

\begin{eqnarray}
y_b(t) & = y_ h(t+ \varphi) + y_ {\theta}(t + \phi)
\end{eqnarray}

Spacing $\mathcal{S}$ is the distance between the TE of the front foil and the leading edge (LE) of the back foil towards the streamwise direction. Therefore the chord normalised spacing $\mathcal{S}_C$ is defined as:

\begin{equation}
\mathcal{S}_\mathcal{C} = \frac{\mathcal{S}}{\mathcal{C}}
\end{equation}

Here $\mathcal{S}_C=2$ to allow comparison with relevant studies of the same $St_A$ \citep{Muscutt2017}. Moreover, $\phi=0^o$ as preliminary simulations found that it maximizes thrust augmentation for the chosen $\mathcal{S}_\mathcal{C}$ and $St_A$ in 2D, the details
of which can be found in Appendix \ref{appendix:a}.

\subsection{Performance Metrics}
Within a flapping cycle, the flipper experiences the time-dependent thrust $F_{X}(t)$, the side force (lift/downforce) $F_{Y}(t)$ and moment $M_{Z}(t)$ around $\mathcal{P}$. In this study we focus on the thrust generation capacity of the fore and hind flipper, characterised by the thrust coefficient:
\begin{equation}
C_{T} = \frac{F_{X}}{\frac{1}{2}\rho U_\infty ^2 G} 
\end{equation}
where $G= \mathcal{C} \cdot (W-\mathcal{C}) + 0.25 \pi \mathcal{C}^2$ is the planform area. Cycle averaged quantities are presented with a tilde ($~\widetilde{}~$) to distinguish them from their instantaneous counterparts. Furthermore, the two dimensional thrust coefficient (where $G$ is replaced by $\mathcal{C}$) is distinguished by the use of the subscript $t$ instead of $T$.

Another important parameter is the propulsive (hydrodynamic) efficiency ($\eta$) of each flipper. This is simply the ratio between the power of the generated thrust and the power imparted to the flipper so that it overcomes the loads imposed by the fluid:
\begin{equation}
  \eta =\frac{T U_\infty}{P}
\end{equation}
where $T$ is the thrust and $P$ the input power defined as:
\begin{equation}
    P(t)=F_{Y}(t)\frac{dy_ h(t)}{dt}+ M_Z(t)\frac{d\theta(t)}{dt} 
\end{equation}

To compare the performance of the single and tandem configurations we normalize the above values by the equivalent parameters of the single (front) flipper:
\begin{equation}
C^*_{T,b} = \frac{\tilde{C_{T,b}}}{\tilde{C_{T,f}}} ~~,~~\eta^*_{b} = \frac{\eta_{b}}{\eta_{f}}
\end{equation}
where $^*$ denotes relative terms. Previous studies suggest little to no alteration of the front foil´s loads and efficiency by the presence of the hind for $S_\mathcal{C} \ge 1$ \citep{Muscutt2017}. Thus, any normalised parameters presented here are associated with the back flipper.
\subsection{Computational Method}
The CFD solver selected for this work can simulate complex geometries and moving boundaries for a variety of Reynolds numbers in 2D and 3D domains, by utilizing the boundary data immersion method (BDIM). BDIM solves a combined set of analytic meta-equations for an immersed solid and its ambient flow, achieving a smoothed interface via an integral kernel function \citep{Weymouth2011}. The technique shows a quadratic convergence and has been validated for flapping foil applications at a wide range of kinematics \citep{Maertens2015,Polet2015}.

The mesh configuration is formed by a rectangular Cartesian grid with a dense uniform domain around the body and near wake while exponential grid stretching is used for the far-field. The boundary conditions consist of a uniform inflow, zero-gradient out flow and free-slip conditions on the upper and lower boundaries. Furthermore no slip boundary conditions are used on the oscillating foil and symmetric conditions are enforced towards the spanwise direction. Mesh density is expressed in terms of grid points per chord. A
uniform grid of $\Delta x = \Delta y = \Delta z = C/64$ is used for this study, yielding relatively fast results while the standard deviation of the estimated thrust is $\leq 8\%$ of the simulations with four times the resolution in both 2D and 3D \citep{Andhini2021}.

\section{Results and Discussion}
\subsection{AR effect on the single flipper}
 
\begin{figure}
  \centerline{\includegraphics[width =5 in]{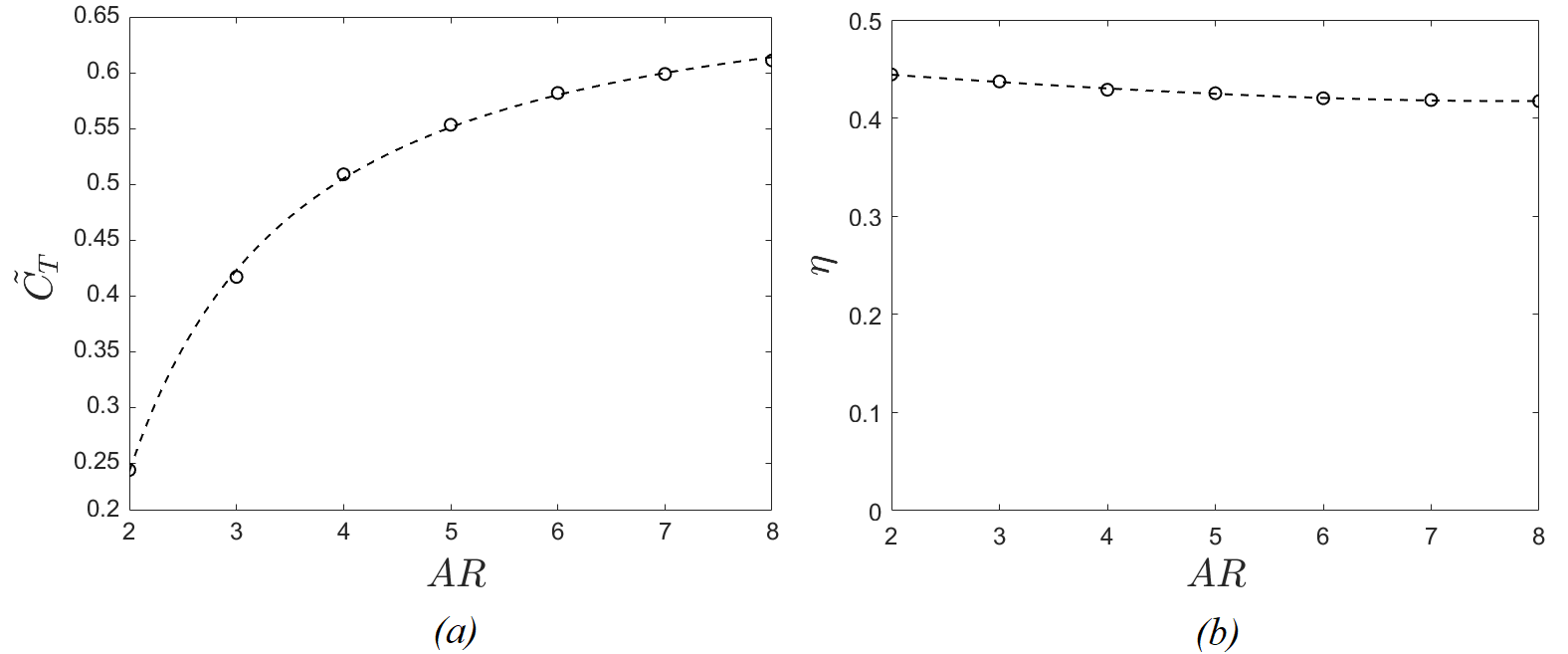}}
  \caption{The impact of AR on the (a) thrust coefficient  and (b) efficiency of the single flipper, undergoing heave-to-pitch coupling. Simulation points are characterised by $\circ$ while the best fit curve is depicted via a dashed line.}\label{single}
\end{figure}

The performance of the single flipper at varying AR can be seen in figure \ref{single}. Elongation leads to a sharp increase of the thrust coefficient until $AR \sim 4$ where the curve starts to asymtote for higher AR, where $\frac{\delta \tilde{C_T}}{\delta AR} \leq 3\%$ (beyond AR > 6). This is still far away from the two dimensional value of $\tilde{C_{t,f}} =0.675$ (see Appendix \ref{appendix:a}). Unlike $\tilde{C_T}$, the propulsive efficiency seems almost insensitive to slenderness possibly due to the use of optimal kinematic parameters, with a negligible decline observed for $AR$ between 2 and 4. 

\begin{figure}
  \centerline{\includegraphics[width =4.5 in]{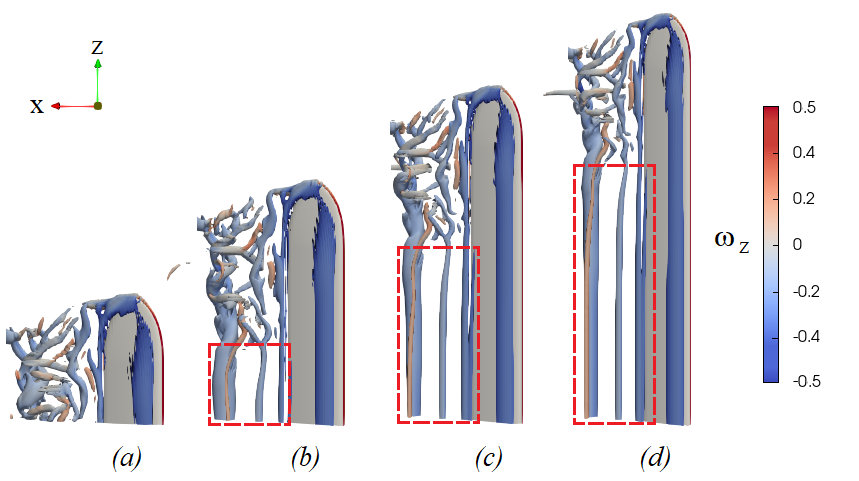}}
  \caption{Snapshots of normalised vorticity at $t/T = 1$ where $T=1/f$, for single flappers of (a) $AR=2$, (b) $AR=4$, (c) $AR=6$ and (d) $AR=8$. The direction of the free-stream flow $U_{\infty}$ is right to left. Areas of undisturbed, two dimensional wake are characterised by rectangles of red dashed lines.}\label{singViz}
\end{figure}

A qualitative comparison of the flow field around different AR´s can be seen in figure \ref{singViz}. The flapping motion generates a cylindrical vortex across the leading edge (LEV) of the flipper which appears almost undisturbed by the tip effects. Turbulent structures can be seen only downstream of the elliptical edge, indicating local breakdown of the trailing edge vortex (TEV). This breakdown propagates towards the root covering a distance of $2.5~\mathcal{C}$ which remains constant, regardless of flipper AR. Consequently, an increasing AR allows the formation of elongated, undisturbed TEVs leading to quasi-two-dimensional wake.

\begin{figure}
  \centerline{\includegraphics[width =5 in]{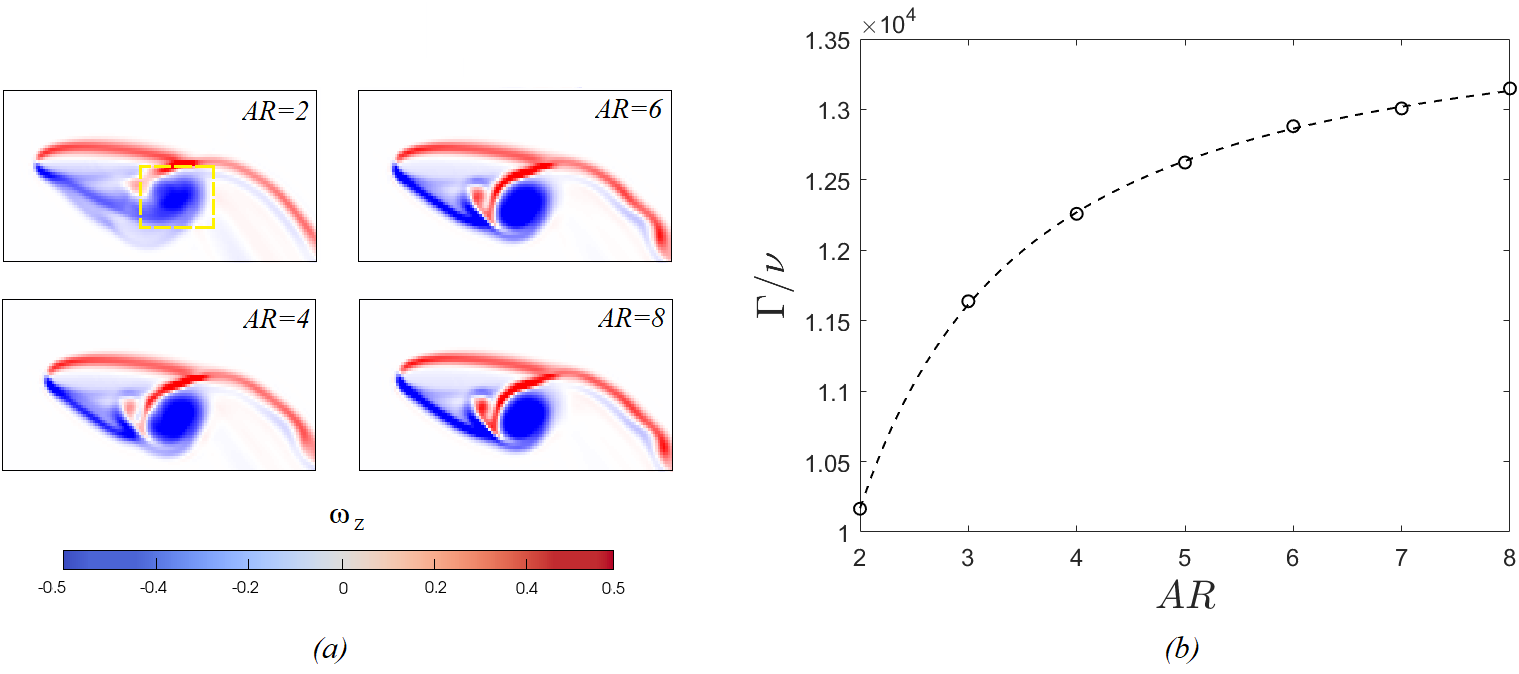}}
  \caption{(a) Spanwise averaged vorticity for single flippers at $t/T = 0.5$ where the TEV is enclosed by a box of yellow dashed line for $AR=2$. (b) The resultant circulation over kinematic viscosity ($\nu$) of the TEV, calculated at this instance for all AR´s of this study. }\label{singGamma}
\end{figure}

In order to quantify the above observations, we examine changes in the strength of the downstream wake, for varying levels of planform slenderness. This can be achieved by calculating the circulation ($\Gamma$) of a TEV at a chosen instant of the flapping cycle, from a spanwise-averaged flow field. Variations along the span will be absorbed in this spanwise-averaging process leading to increased circulation for only the most coherent TEVs. Details of the  procedure can be found in Appendix \ref{appendix:b}. As shown in figure \ref{singGamma}a, the TEV appears to become more compact for higher AR, causing $\Gamma$ to saturate at a constant value (see figure \ref{singGamma}b). This in turn, leads to a constant velocity surplus across the flipper span which is reflected in behaviour of the thrust in figure \ref{single}. 

Here we should note that, similar $ \tilde{C_T}-AR$ relationships have been reported by \cite{Shao2010} despite the latter´s different planform geometry (no wingtip) and significantly lower $Re_\mathcal{C}$. As both studies utilize heave-to-pitch coupling at $\psi=90^o$, a similar dynamic separation should be expected at least in 2D. Moreover, the chosen kinematics are essentially two dimensional ($\frac{\delta y_f}{\delta z} =\frac{\delta y_f}{\delta z}=0$), further reducing the importance of the spanwise geometry, although they still cause minor discrepancies between the two studies e.g. in $\tilde{C_T}$. This, in turn, can support the predominance of medium  slenderness ($AR \sim [4, 6]$ using our paper´s definition) caudal fins observed in cetaceans \citep{Ayancik2020} where the same kinematics are used. Indeed, fins of too low AR would reduce the propulsive capacity of the species, yet much larger ones would be structurally demanding without offering any significant hydrodynamic advantage.  It should be noted, however, that these animals demonstrate a plethora of wingtip geometries, combined with at least some level of flexibility \citep{Fish1999}. Thus, although a deeper analysis in the area is required, this topic is beyond the focus of the present work. 

\subsection{AR effect on the tandem configuration}
\begin{figure}
  \centerline{\includegraphics[width =5.5 in]{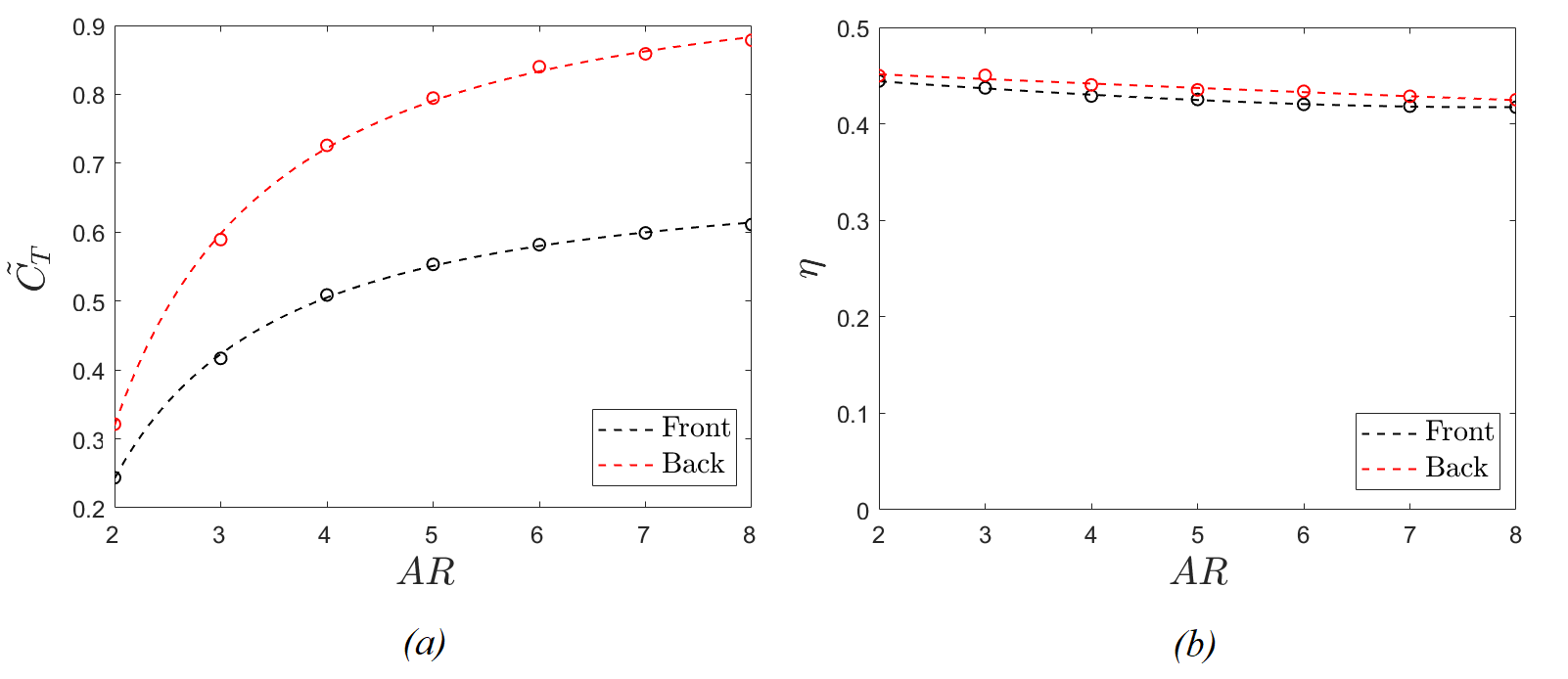}}
  \caption{The impact of AR in terms of (a) thrust  coefficient and (b) efficiency, on the fore and hind flippers of a tandem configuration, undergoing heave-to-pitch coupling at $\phi=0^{\circ}$ and $\mathcal{S}_{\mathcal{C}}=2$. Simulation points are characterised by $\circ$ while the best fit curve is depicted via a dashed line.}\label{tandem}
\end{figure}

The addition of another upstream oscillating body alters the flow field and determines the propulsive characteristics of the downstream foil. This is made clear in figure \ref{tandem}a where the rear flipper demonstrates significantly higher $\tilde{C_T}$ than the front, due to wake recapture. Once again, elongation has a greater impact on low AR´s but $\tilde{C_T}$ appears to stabilize at a rate that is marginally slower than that of a single foil (or that of the front foil). This suggest that the wake recapture tends to benefit over a larger range of ARs compared to performance improvement of a single foil. On the other hand, the efficiency shows a negligible improvement while following the same trend as for the front foil (see figure \ref{tandem}b). This is not surprising since the optimal $\phi$ for thrust augmentation often differs from the one of improved efficiency \citep{Muscutt2017}.

\begin{figure}
  \centerline{\includegraphics[width =5.5 in]{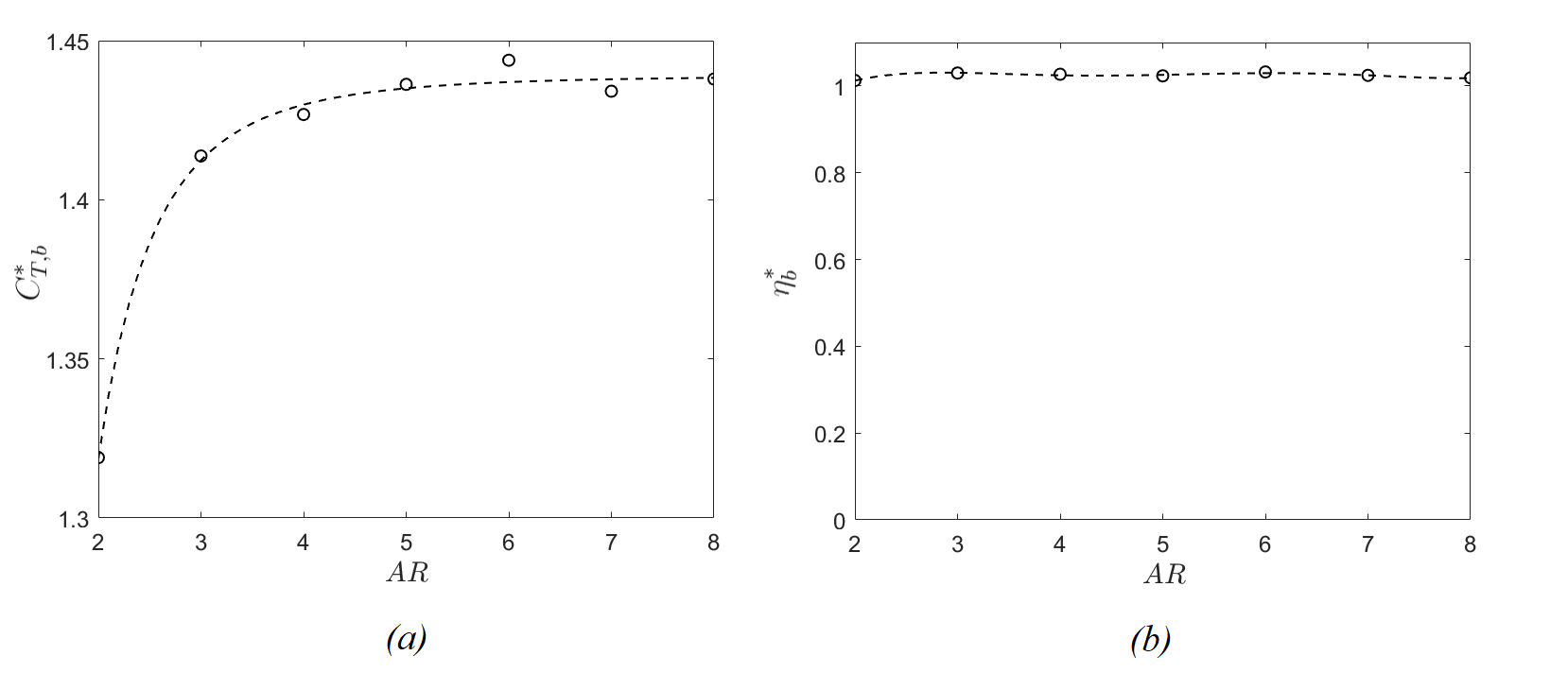}}
  \caption{The impact of AR, in terms of relative (a) thrust and (b) efficiency augmentation, on the hind flipper of the tandem configuration, undergoing heave-to-pitch coupling at $\phi=0^{\circ}$ and $\mathcal{S}_{\mathcal{C}}=2$. Simulation points are depicted as $\circ$ while the dashed lines represent the best fit curves. }\label{relative}
\end{figure}

In order to investigate the relative augmentation of thrust for the back foil in more detail, we examine $\tilde{C_{T,b}}^*$ (Which is the ratio of thrust of the back foil to that of the front foil) in figure \ref{relative}a. It can be seen that there is a sharp increase in this ratio for $AR \sim [2, 4]$ (from $\tilde{C_{T,b}}^* =1.3$ to $\tilde{C_{T,b}}^* =1.45$) and the ratio seems to level out around AR = 4 (at $\tilde{C_{T,b}}^* \sim 1.42$) which remains approximately constant beyond this aspect ratio. This suggests that the rate of increase in thrust for the front and back foils essentially follow each other proportionately. Consequently, there are no further benefits beyond AR = 4 in terms of relative augmentation, although there is still a benefit in the overall thrust produced by the pair of flippers.

Interestingly, figure \ref{relative}b shows that relative efficiency ${\eta_b}^*$ (which is the ratio of efficiency of back foil to the front foil) remains practically unchanged, showing very minor gains of about $2.4\%$ throughout the entire range of AR (see figure \ref{relative}b). This is probably related to the flipper kinematics being already optimized for maximum efficiency, in conjunction with the, previously mentioned, use of a thrust-specific $\phi$. Thus, it may be possible to alter the kinematics and/or the planform of the flippers to make further gains in this area. However, this is beyond the scope of the current study. 

\begin{figure}
  \centerline{\includegraphics[width =6 in]{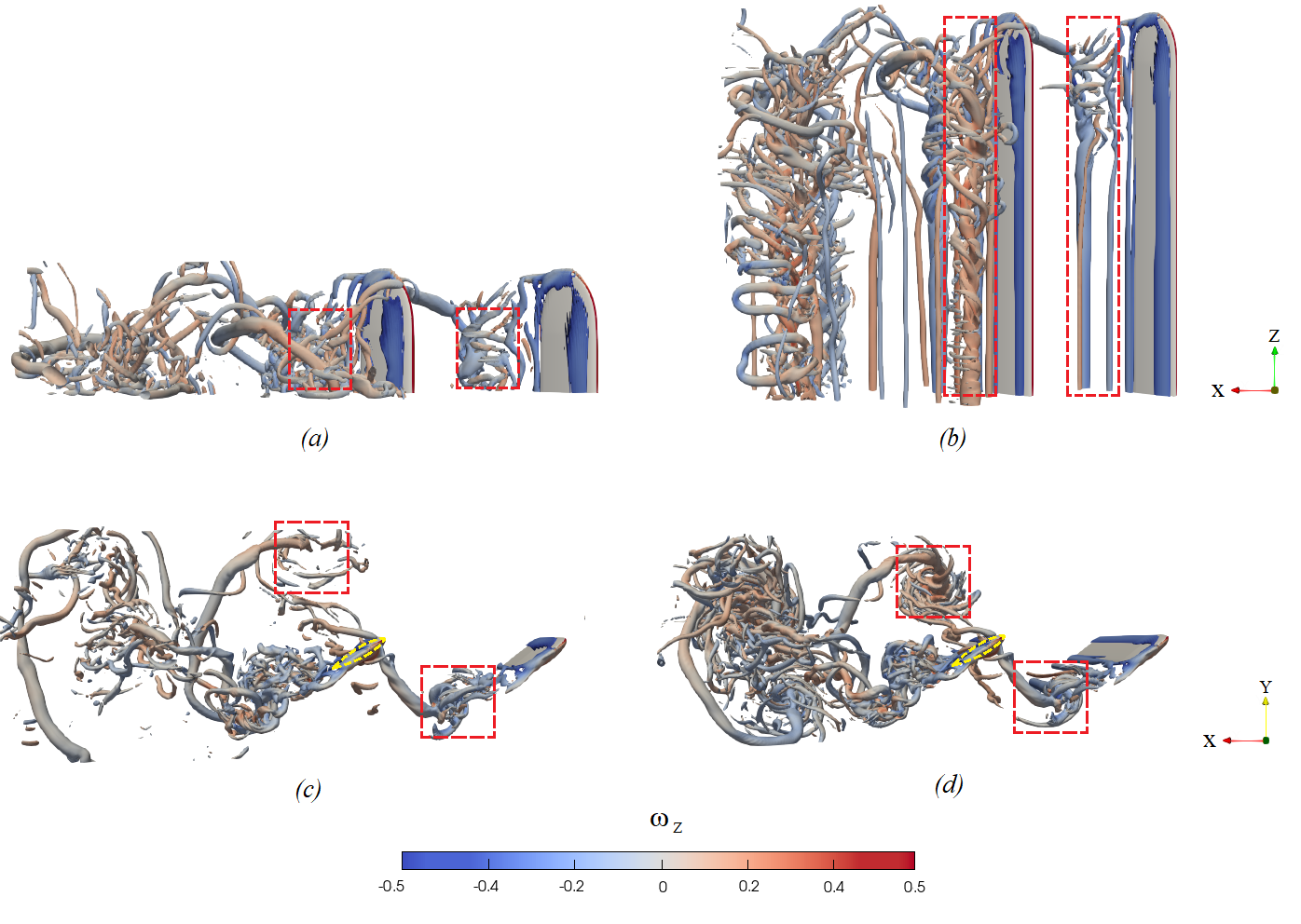}}
  \caption{Snapshots of normalised vorticity at $t/T=1$ for tandem configurations.  A top view comparison shows that the wake of $AR=2$ (a) suffers significantly from vortex breakdown  while the wake of $AR=8$ (c) remains mostly unaffected. This is more evident at a side view although the back foil (yellow dashed line) of both $AR=2$ (c) and  $AR=8$ (d) manages to weave through the incoming vortex pair (boxes of red dashed line) of the front flipper, due to proper $\phi$ adjustment.}\label{tandemViz}
\end{figure}

The above observations are obviously related to the flow field development between the two foils. This is evident in figure \ref{tandemViz}, where the wakes of tandem arrangements for $AR=2$ and $AR=8$ are compared (animations of these test cases can be found in the supplementary material). As mentioned previously, wingtip effects are proportionally higher in the wake of $AR=2$ compared to $AR=8$ (figures \ref{tandemViz}a and \ref{tandemViz}b) resulting in the break-up of TEV shed from the front and the back foils (figures \ref{tandemViz}c and \ref{tandemViz}d). Specifically, the break-up of the TEV from the front foil means that the back foil does not experience a coherent wake across its span. Therefore, any additional benefit that could come from wake-vortex capture is not significant. This also limits paired effects of vortices in the wake of the rear-foil. The figure shows that the distance between shed-vortices of the two foils appears to decrease with increasing aspect ratio and this is consistent observations in previous studies \citep{Dong2006, Shao2010}. This also suggests that low AR foils enable a greater range of phase-difference in the kinematics between the two foils since it is easier for the back foil to weave between the incoming vortices, which is a key characteristic of wake recapture \citep{Muscutt2017}. Finally, as AR increases, the incoming wake for the back foil takes a  quasi-2D form (except at around the tip) and relative augmentation begins to stagnate since no additional benefit is extracted. 

\begin{figure}
  \centerline{\includegraphics[width =5 in]{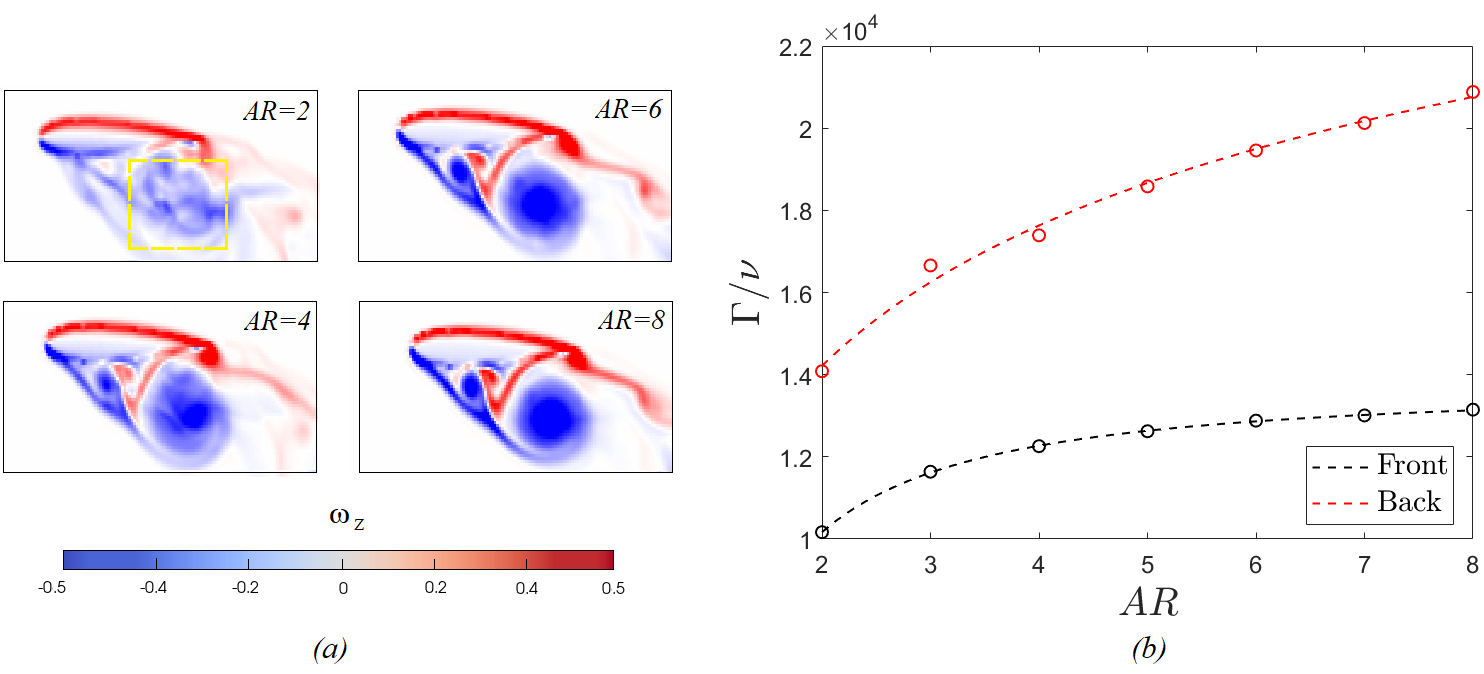}}
  \caption{(a) Spanwise averaged vorticity for back flippers of a tandem configuration, at $t/T = 0.5$ where the TEV is enclosed in a box of yellow dashed line for $AR=2$. (b) The resultant $\Gamma / \nu$ calculated at this instance, for the TEV of both front and back foils with $AR  \sim [2,8]$. }\label{tanGamma}
\end{figure}

These findings can be further quantified by computing the spanwise-averaged circulation of the TEV or the back flipper for different aspect ratios. As shown in figure \ref{tanGamma}a, wake recapture allows the formation of a noticeably larger and stronger TEV compared with those shed by the front foil. However, its compactness/coherence is more dependent on flipper elongation, which alters the TEV circulation with AR (see figure \ref{tanGamma}b). Note that the circulation is computed based on a given box size and further information on the effect of box size on the computation of this circulation is in Appendix \ref{appendix:b}. For low AR, the vortex appears to be diffused due to interactions between the main TEV and the tip. As AR increases, the spanwise-averaged TEV becomes more coherent and its circulation rises (for a given box size). However, this value starts to level off at higher AR, following the same trend as the thrust coefficient. It can also be seen that the circulation values in the back foil are significantly higher than that of the front foil, which is also consistent with the thrust results. 

From an application point of view, the above findings suggest that a set of high AR flippers would be preferable for steady cruising, while lower AR permit a more agile behaviour at the cost of speed. Moreover, tandem arrangements of $AR \sim [4-6]$ may be a prudent compromise between augmentation benefits and mechanical behaviour, since the relative thrust enhancement has reached a saturated state while the size is still small enough to withstand the large unsteady loads. Finally,we speculate that it might be beneficial for the rear foil to be at slightly lower AR than the front in order to avoid the tip vortex effect that could reduce the relative augmentation. 

\vspace*{-5pt}
\section{Conclusions}
The propulsive characteristics of single and tandem flapping foils were examined numerically under a heave-to-pitch coupling motion, for seven flipper sets of $AR \sim [2-8]$ of rectangular flippers with elliptical tip at $Re_{\mathcal{C}}=8500$. Each set had the same AR and the test were conducted for the fixed combination $St_A =0.4 - \mathcal{S}_C=2$ at $\phi =0^o$ which was found to optimise wake recapture in 2D. 

Our analysis shows that flipper elongation has a positive impact on the thrust coefficients of both single and tandem configurations at low AR´s but this effect weakens as we move towards higher AR. More specifically, an increasing AR benefits the wake recapture of the tandem configuration, which results in a slower convergence of the back flipper´s thrust coefficient. On the other hand the efficiency remains virtually unaffected due to the flippers´ optimal kinematics and the thrust-targeting $\phi$. 

Fundamentally, the above are related to the enhanced strength and cohesion of the TEV´s pair shed at each stroke. In particular, snapshots of instantaneous vorticity show that 3D effects have a localized behaviour around the wing tip of the front foil which remains constant throughout the range of AR´s. This affects spanwise-averaged TEV circulation, which increases with elongation but eventually saturates, so that the benefits diminish towards two dimensional concepts ($AR \sim \infty$). Low AR´s, however, lead to weaker TEV´s which move away from the centerline and decay faster in the streamwise direction. This is vital for the hind flipper as a stronger incoming wake will induce higher acceleration of the surrounding flow while a wider, weaker vortex pair can enable a greater range of $\phi$ since weaving within the vortices becomes easier. Consequently, wake recapture leads to an augmented $\Gamma$ for the TEV of the rear foil, combined with a higher sensitivity towards flipper slenderness. This results in a comparatively sharper circulation increase and a slower convergence, in a similar fashion to the $\tilde{C_T}$ results.

This study provides evidence of the AR impact on wake recapture under kinematics commonly used in the natural world. In addition, our findings provide some design rules for more versatile bio-inspired systems by revealing the hydrodynamic benefits and limitations of the single and tandem flipper arrangements. It should be noted that, the singular effect of aspect ratio has been addressed here for a simple planform. However, natural systems can achieve a wide range of planforms for the same aspect ratio and it is possible to design even more through engineering tools. Therefore, the effects of planform shape can be further explored for a given aspect ratio that is most suitable for tandem foils. 

\begin{figure}
  \centerline{\includegraphics[width =3 in]{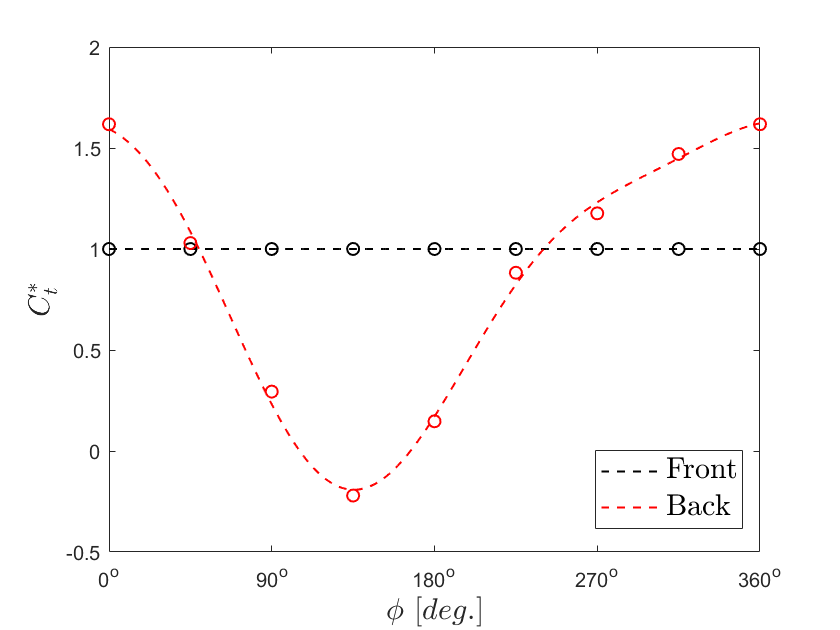}}
  \caption{Impact of $\phi$ on the two dimensional wake recapture, expressed via the relative thrust augmentation of the two foils. Here $C^*_{T,f} =1$ since the front foil experiences no flow field changes, which coincides with $\tilde{C_{t,f}} \sim 0.675$. Simulation points are depicted as $\circ$ while the dashed lines represent the best fit curves.}\label{2D}
\end{figure}
\appendix
\section*{APPENDIX}
\section{Phase Optimisation} \label{appendix:a}
To set our reference test case in terms of maximum thrust augmentation, a preliminary study was conducted in 2D, evaluating the phasing $\phi$ of the tandem configuration for the chosen spacing, kinematic parameters and ambient conditions. Tandem foil simulations were performed, starting from $\phi =0^o$ and progressing at increments of $\Delta\phi=45^o$ until $\phi =315^o$, while the single foil was found to produce $\tilde{C_{t,f}} \sim 0.675$. Figure \ref{2D}  shows that the modification of the hind foil´s thrust due to interacting with the incoming wake, follows a cosine-like curve with respect to the phase lag, as shown in similar studies \citep{MuscuttEX2017}. Clearly, optimal $C^*_{T,b}$ is found for $\phi =0^o$ and therefore it is chosen for all the simulations presented in the current study. 

\begin{figure}
  \centerline{\includegraphics[width =5 in]{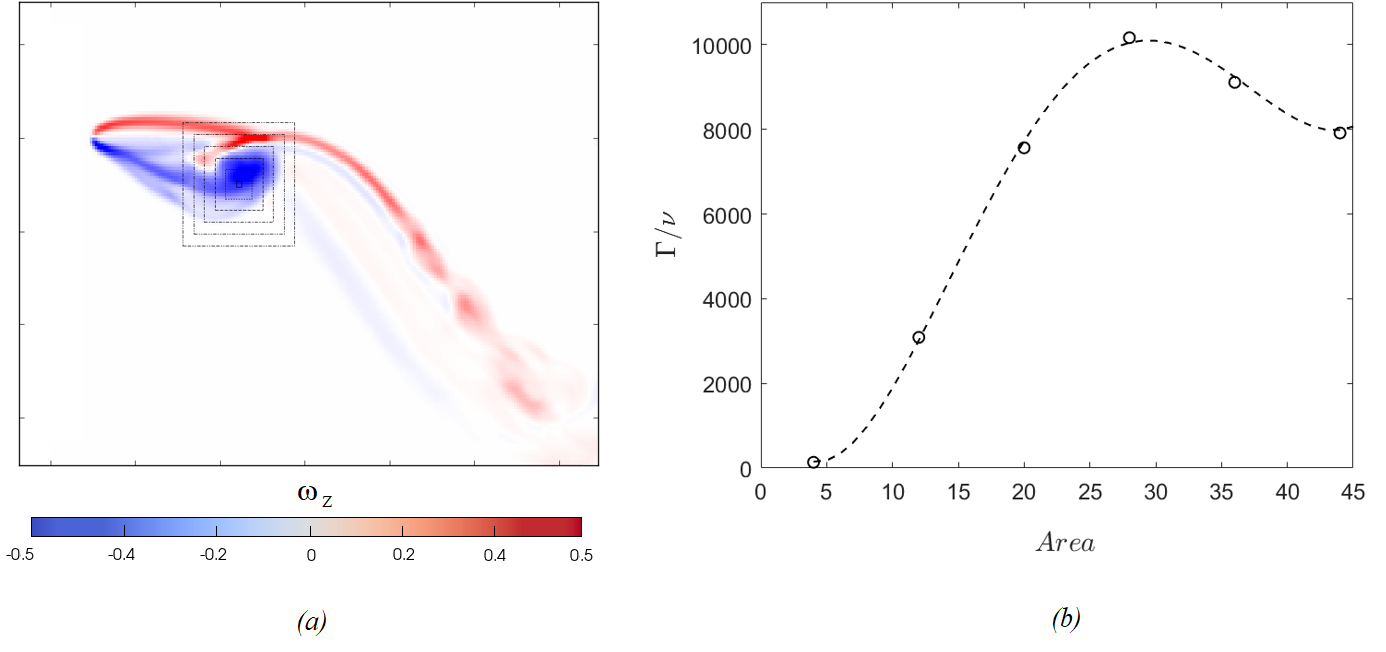}}
  \caption{Sensitivity analysis of the integration area, used to calculate the circulation of the front foil´s TEV. Here the vorticity is first spanwise averaged, at $t/T= 0.5$ as shown in (a) for $AR=2$. Then it is integrated within boxes of increasing size until circulation values begin to drop (b). }\label{fig10}
\end{figure}

\section{Sensitivity Analysis of the TEV Circulation } \label{appendix:b}
In this study, circulation is calculated via the integration of vorticity over a rectangular cell (see figure \ref{fig10}a). As we focus on the TEV analysis, the size and location of this area should be optimized to enclose the exact size of the vortex while minimizing ambient interference. Therefore, after visually choosing an initial location, we gradually increase the area of integration until the overall circulation begins to drop (see figure \ref{fig10}b). Having finalized the size of the box, we re-evaluate its location by moving its center towards the y and x axis until a position of maximum circulation is identified. Due to the shape/size of the TEV for various ARs, this procedure was conducted individually for all single and tandem test cases such that the ``highest'' circulation is obtained for each case. 

\begin{Backmatter}

\paragraph{Acknowledgements}
We would like to thank A.N. Zurman-Nasution and M. Lauber for our fruitful discussions throughout the duration of this project. Furthermore, we would like to thank the IRIDIS High Performance Computing Facility, with its associated support services at the University of Southampton, for their aid towards the completion of our study. 

\paragraph{Funding Statement}
This research was supported financially by the Office of Naval Research award N62909-18-1-2091 and the Engineering and Physical Sciences Research Council doctoral training award [1789955].

\paragraph{Declaration of Interests}
The authors declare no conflict of interest.

\paragraph{Author Contributions}
Conceptualization: N.S.L, G.W. and B.G.
Investigation: N.S.L.
Writing original draft: N.S.L.
Writing review and editing: G.W. and B.G.

\paragraph{Data Availability Statement}
All data supporting this study, including useful supplementary material, is openly available via the University of Southampton repository at (AVAILABLE UPON ACCEPTANCE).

\paragraph{Ethical Standards}
The research meets all ethical guidelines, including
adherence to the legal requirements of the study country.

\bibliographystyle{apalike}
\bibliography{FLO_v1}

\begin{thebibliography}{}

\bibitem[Adams and Fish, 2019]{Adams2019}
Adams, D.~S. and Fish, F.~E. (2019).
\newblock Odontocete peduncle tendons for possible control of fluke orientation
  and flexibility.
\newblock {\em Journal of morphology}, 280(9):1323--1331.

\bibitem[Alexander, 1984]{Alexander1984}
Alexander, D.~E. (1984).
\newblock Unusual phase relationships between the forewings and hindwings in
  flying dragonflies.
\newblock {\em Journal of Experimental Biology}, 109(1):379--383.

\bibitem[Arranz et~al., 2020]{Arranz2020}
Arranz, G., Flores, O., and Garcia-Villalba, M. (2020).
\newblock Three-dimensional effects on the aerodynamic performance of flapping
  wings in tandem configuration.
\newblock {\em Journal of Fluids and Structures}, 94:102893.

\bibitem[Ayancik et~al., 2020]{Ayancik2020}
Ayancik, F., Fish, F.~E., and Moored, K.~W. (2020).
\newblock Three-dimensional scaling laws of cetacean propulsion characterize
  the hydrodynamic interplay of flukes' shape and kinematics.
\newblock {\em Journal of the Royal Society Interface}, 17(163):20190655.

\bibitem[Azuma, 1992]{Azuma1992}
Azuma, A. (1992).
\newblock Flight by beating.
\newblock In {\em The Biokinetics of Flying and Swimming}, pages 77--154.
  Springer.

\bibitem[Broering and Lian, 2012]{Broering2012}
Broering, T.~M. and Lian, Y.-S. (2012).
\newblock The effect of phase angle and wing spacing on tandem flapping wings.
\newblock {\em Acta Mechanica Sinica}, 28(6):1557--1571.

\bibitem[Carpenter et~al., 2010]{Carpenter2010}
Carpenter, K., Sanders, F., Reed, B., Reed, J., and Larson, P. (2010).
\newblock Plesiosaur swimming as interpreted from skeletal analysis and
  experimental results.
\newblock {\em Transactions of the Kansas Academy of Science}, 113(1/2):1--34.

\bibitem[Cimarelli et~al., 2021]{Cimarelli2021}
Cimarelli, A., Franciolini, M., and Crivellini, A. (2021).
\newblock On the kinematics and dynamics parameters governing the flow in
  oscillating foils.
\newblock {\em Journal of Fluids and Structures}, 101:103220.

\bibitem[Combes and Daniel, 2001]{Combes2001}
Combes, S. and Daniel, T. (2001).
\newblock Shape, flapping and flexion: wing and fin design for forward flight.
\newblock {\em Journal of Experimental Biology}, 204(12):2073--2085.

\bibitem[Dagenais and Aegerter, 2020]{Dagenais2020}
Dagenais, P. and Aegerter, C.~M. (2020).
\newblock How shape and flapping rate affect the distribution of fluid forces
  on flexible hydrofoils.
\newblock {\em Journal of Fluid Mechanics}, 901.

\bibitem[Dewey et~al., 2013]{Dewey2013}
Dewey, P.~A., Boschitsch, B.~M., Moored, K.~W., Stone, H.~A., and Smits, A.~J.
  (2013).
\newblock Scaling laws for the thrust production of flexible pitching panels.
\newblock {\em Journal of Fluid Mechanics}, 732:29.

\bibitem[Domenici, 2010]{Domenici2010}
Domenici, P. (2010).
\newblock {\em Fish locomotion: an eco-ethological perspective}.
\newblock CRC Press.

\bibitem[Dong et~al., 2006]{Dong2006}
Dong, H., Mittal, R., and Najjar, F. (2006).
\newblock Wake topology and hydrodynamic performance of low-aspect-ratio
  flapping foils.
\newblock {\em Journal of Fluid Mechanics}, 566:309.

\bibitem[Ellington, 1984]{Ellington1984}
Ellington, C.~P. (1984).
\newblock The aerodynamics of hovering insect flight. ii. morphological
  parameters.
\newblock {\em Philosophical Transactions of the Royal Society of London. B,
  Biological Sciences}, 305(1122):17--40.

\bibitem[Fernandez-Feria and Alaminos-Quesada, 2021]{Fernandez2021}
Fernandez-Feria, R. and Alaminos-Quesada, J. (2021).
\newblock Analytical results for the propulsion performance of a flexible foil
  with prescribed pitching and heaving motions and passive small deflection.
\newblock {\em Journal of Fluid Mechanics}, 910.

\bibitem[Fish and Rohr, 1999]{Fish1999}
Fish, F.~E. and Rohr, J. (1999).
\newblock Review of dolphin hydrodynamics and swimming performance.

\bibitem[Flammang and Lauder, 2009]{Flammang2009}
Flammang, B.~E. and Lauder, G.~V. (2009).
\newblock Caudal fin shape modulation and control during acceleration, braking
  and backing maneuvers in bluegill sunfish, lepomis macrochirus.
\newblock {\em Journal of Experimental Biology}, 212(2):277--286.

\bibitem[Gough et~al., 2018]{Gough2018}
Gough, W.~T., Fish, F.~E., Wainwright, D.~K., and Bart-Smith, H. (2018).
\newblock Morphology of the core fibrous layer of the cetacean tail fluke.
\newblock {\em Journal of morphology}, 279(6):757--765.

\bibitem[Green and Smits, 2008]{Green2008}
Green, M.~A. and Smits, A.~J. (2008).
\newblock Effects of three-dimensionality on thrust production by a pitching
  panel.
\newblock {\em Journal of fluid mechanics}, 615:211--220.

\bibitem[Guglielmini and Blondeaux, 2004]{Guglielmini2004}
Guglielmini, L. and Blondeaux, P. (2004).
\newblock Propulsive efficiency of oscillating foils.
\newblock {\em European Journal of Mechanics-B/Fluids}, 23(2):255--278.

\bibitem[Hammer et~al., 2021]{Hammer2021}
Hammer, P.~R., Garmann, D.~J., and Visbal, M. (2021).
\newblock Aspect ratio effect on finite wing dynamic stall.
\newblock In {\em AIAA Scitech 2021 Forum}, page 1089.

\bibitem[Han et~al., 2020]{Han2020}
Han, P., Wang, J., Fish, F.~E., and Dong, H. (2020).
\newblock Kinematics and hydrodynamics of a dolphin in forward swimming.
\newblock In {\em AIAA AVIATION 2020 FORUM}, page 3015.

\bibitem[Hawthorne et~al., 2019]{Hawthorne2019}
Hawthorne, M., McMenamin, M., and De~la Salle, P. (2019).
\newblock How plesiosaurs swam: New insights into their underwater flight using
  “ava”, a virtual pliosaur.

\bibitem[Khalid et~al., 2021]{Khalid2021}
Khalid, M. S.~U., Wang, J., Akhtar, I., Dong, H., Liu, M., and Hemmati, A.
  (2021).
\newblock Why do anguilliform swimmers perform undulation with wavelengths
  shorter than their bodylengths?
\newblock {\em Physics of Fluids}, 33(3):031911.

\bibitem[Kim et~al., 2013]{Kim2013}
Kim, D., Hussain, F., and Gharib, M. (2013).
\newblock Vortex dynamics of clapping plates.
\newblock {\em Journal of Fluid Mechanics}, 714:5--23.

\bibitem[Koochesfahani, 1989]{Koochesfahani1989}
Koochesfahani, M. (1989).
\newblock Vortical patterns in the wake of an oscillating airfoil.
\newblock {\em AIAA journal}, 27(9):1200--1205.

\bibitem[Lagopoulos et~al., 2019]{Lagopoulos2019}
Lagopoulos, N.~S., Weymouth, G.~D., and Ganapathisubramani, B. (2019).
\newblock Universal scaling law for drag-to-thrust wake transition in flapping
  foils.
\newblock {\em Journal of Fluid Mechanics}, 872.

\bibitem[Lee et~al., 2017]{Lee2017}
Lee, J., Park, Y.-J., Cho, K.-J., Kim, D., and Kim, H.-Y. (2017).
\newblock Hydrodynamic advantages of a low aspect-ratio flapping foil.
\newblock {\em Journal of Fluids and Structures}, 71:70--77.

\bibitem[Li et~al., 2018]{Li2018}
Li, Y., Pan, D., Zhao, Q., Ma, Z., and Wang, X. (2018).
\newblock Hydrodynamic performance of an autonomous underwater glider with a
  pair of bioinspired hydro wings--a numerical investigation.
\newblock {\em Ocean Engineering}, 163:51--57.

\bibitem[Licht et~al., 2004]{Licht2004}
Licht, S., Polidoro, V., Flores, M., Hover, F.~S., and Triantafyllou, M.~S.
  (2004).
\newblock Design and projected performance of a flapping foil auv.
\newblock {\em IEEE Journal of oceanic engineering}, 29(3):786--794.

\bibitem[Long et~al., 2006]{Long2006}
Long, J.~H., Schumacher, J., Livingston, N., and Kemp, M. (2006).
\newblock Four flippers or two? tetrapodal swimming with an aquatic robot.
\newblock {\em Bioinspiration and Biomimetics}, 1(1):20.

\bibitem[Maertens and Weymouth, 2015]{Maertens2015}
Maertens, A. and Weymouth, G. (2015).
\newblock Accurate cartesian-grid simulations of near-body flows at
  intermediate reynolds numbers.
\newblock {\em Computer Methods in Applied Mechanics and Engineering}, 283:106
  -- 129.

\bibitem[Mittal, 2004]{Mittal2004}
Mittal, R. (2004).
\newblock Computational modeling in biohydrodynamics: Trends, challenges, and
  recent advances.
\newblock {\em IEEE Journal of Oceanic Engineering}, 29(3):595--604.

\bibitem[Muscutt et~al., 2017a]{MuscuttEX2017}
Muscutt, L.~E., Dyke, G., Weymouth, G.~D., Naish, D., Palmer, C., and
  Ganapathisubramani, B. (2017a).
\newblock The four-flipper swimming method of plesiosaurs enabled efficient and
  effective locomotion.
\newblock {\em Proceedings of the Royal Society B: Biological Sciences},
  284(1861):20170951.

\bibitem[Muscutt et~al., 2017b]{Muscutt2017}
Muscutt, L.~E., Weymouth, G.~D., and Ganapathisubramani, B. (2017b).
\newblock Performance augmentation mechanism of in-line tandem flapping foils.
\newblock {\em Journal of Fluid Mechanics}, 827:484–505.

\bibitem[Pedro et~al., 2003]{Pedro2003}
Pedro, G., Suleman, A., and Djilali, N. (2003).
\newblock A numerical study of the propulsive efficiency of a flapping
  hydrofoil.
\newblock {\em International journal for numerical methods in fluids},
  42(5):493--526.

\bibitem[Platzer and Jones, 2008]{Platzer2008}
Platzer, M. and Jones, K. (2008).
\newblock Flapping wing aerodynamics-progress and challenges.
\newblock In {\em 44th AIAA Aerospace Sciences Meeting and Exhibit}, page 500.

\bibitem[Polet et~al., 2015]{Polet2015}
Polet, D., Rival, D., and Weymouth, G. (2015).
\newblock Unsteady dynamics of rapid perching manoeuvres.
\newblock {\em Journal of Fluid Mechanics}, 767:323–341.

\bibitem[Robinson and JA, 1975]{Robinson1975}
Robinson, J.~A. and JA, R. (1975).
\newblock The locomotion of plesiosaurs.

\bibitem[Shao et~al., 2010]{Shao2010}
Shao, X.-m., Pan, D.-y., Deng, J., and Yu, Z.-s. (2010).
\newblock Numerical studies on the propulsion and wake structures of
  finite-span flapping wings with different aspect ratios.
\newblock {\em Journal of Hydrodynamics}, 22(2):147--154.

\bibitem[Shi et~al., 2020]{Shi2020}
Shi, G., Xiao, Q., and Zhu, Q. (2020).
\newblock Effects of time-varying flexibility on the propulsion performance of
  a flapping foil.
\newblock {\em Physics of Fluids}, 32(12):121904.

\bibitem[Thomas et~al., 2004]{Thomas2004}
Thomas, A. L.~R., Taylor, G.~K., Srygley, R.~B., Nudds, R.~L., and Bomphrey,
  R.~J. (2004).
\newblock Dragonfly flight: free-flight and tethered flow visualizations reveal
  a diverse array of unsteady lift-generating mechanisms, controlled primarily
  via angle of attack.
\newblock {\em Journal of Experimental Biology}, 207(24):4299--4323.

\bibitem[Triantafyllou et~al., 1993]{Triantafyllou1993}
Triantafyllou, G., Triantafyllou, M., and Grosenbaugh, M. (1993).
\newblock Optimal thrust development in oscillating foils with application to
  fish propulsion.
\newblock {\em Journal of Fluids and Structures}, 7(2):205--224.

\bibitem[Triantafyllou et~al., 1991]{Triantafyllou1991}
Triantafyllou, M., Triantafyllou, G., and Gopalkrishnan, R. (1991).
\newblock Wake mechanics for thrust generation in oscillating foils.
\newblock {\em Physics of Fluids A: Fluid Dynamics}, 3(12):2835–--2837.

\bibitem[Usherwood and Ellington, 2002]{Usherwood2002}
Usherwood, J.~R. and Ellington, C.~P. (2002).
\newblock The aerodynamics of revolving wings ii. propeller force coefficients
  from mayfly to quail.
\newblock {\em Journal of Experimental Biology}, 205(11):1565--1576.

\bibitem[Walker and Westneat, 2002]{Walker2002}
Walker, J.~A. and Westneat, M.~W. (2002).
\newblock Performance limits of labriform propulsion and correlates with fin
  shape and motion.
\newblock {\em Journal of Experimental Biology}, 205(2):177--187.

\bibitem[Weymouth et~al., 2017]{Weymouth2017}
Weymouth, G., Devereux, K., Copsey, N., Muscutt, L., Downes, J., and
  Ganapathisubramani, B. (2017).
\newblock Hydrodynamics of an under-actuated plesiosaur-inspired robot.
\newblock {\em APS}, pages F9--003.

\bibitem[Weymouth and Yue, 2011]{Weymouth2011}
Weymouth, G. and Yue, D. (2011).
\newblock Boundary data immersion method for cartesian-grid simulations of
  fluid-body interaction problems.
\newblock {\em Journal of Computational Physics}, 230(16):6233 -- 6247.

\bibitem[Weymouth, 2016]{Weymouth2016}
Weymouth, G.~D. (2016).
\newblock Biologically inspired force enhancement for maritime propulsion and
  maneuvering.
\newblock {\em arXiv preprint arXiv:1609.06559}.

\bibitem[Yuh, 2000]{Yuh2000}
Yuh, J. (2000).
\newblock Design and control of autonomous underwater robots: A survey.
\newblock {\em Autonomous Robots}, 8(1):7--24.

\bibitem[Zhong et~al., 2021]{Zhong2021}
Zhong, Q., Han, T., Moored, K.~W., and Quinn, D.~B. (2021).
\newblock Aspect ratio affects the equilibrium altitude of near-ground
  swimmers.
\newblock {\em Journal of Fluid Mechanics}, 917.

\bibitem[Zurman-Nasution et~al., 2020]{Zurman2020}
Zurman-Nasution, A., Ganapathisubramani, B., and Weymouth, G. (2020).
\newblock Influence of three-dimensionality on propulsive flapping.
\newblock {\em Journal of Fluid Mechanics}, 886.

\bibitem[Zurman-Nasution et~al., 2021a]{Andhini2021}
Zurman-Nasution, A.~N., Ganapathisubramani, B., and Weymouth, G.~D. (2021a).
\newblock Effects of aspect ratio on rolling and twisting foils.
\newblock {\em Physical Review Fluids}, 6(1):013101.

\bibitem[Zurman-Nasution et~al., 2021b]{Zurman2021}
Zurman-Nasution, A.~N., Ganapathisubramani, B., and Weymouth, G.~D. (2021b).
\newblock Fin sweep angle does not determine flapping propulsive performance.
\newblock {\em Journal of the Royal Society}, 18.

\end{thebibliography}

\end{Backmatter}
\end{document}